\documentstyle[12pt]{article}
\date{}
\begin{document}
\title {\bf Radial pulsation frequencies of slowly rotating neutron stars}
\author{P. K. Sahu and A. R. Prasanna\\
Theory Group, Physical Research Laboratory, Ahmedabad
380 009, India;\\ E-mail: pradip or prasanna@prl.ernet.in.}
\maketitle
\begin{abstract}
We study the radial pulsation frequencies of slowly rotating neutron stars 
in general relativistic formalism using realistic equations of state. It is
found that the pulsation frequencies are always an increasing function of
rotation rate. The increasing rate of frequency depends on the nature
of equations of state.
\end{abstract}
\vskip 0.4in
Cameron \cite{cam65} suggested that the vibration of neutron stars might
excite motions that might have interesting astrophysical applications, which
lead to a series of investigations of the vibrational properties of neutron
stars. The earliest detailed calculations were done by Meltzer and Thorne \cite{mel66}
and Thorne \cite{tho69}, where they investigated the radial as well as nonradial 
oscillations using available equation of state, such as the Harrison -
Walker - Wheeler equation of state. These and other early studies by 
Wheeler \cite{whe66}, Chau \cite{cha67} and Occhionero \cite{occ68} 
indicated that the majority of the 
fundamental mode radial oscillation periods for the neutron stars would lie
in the vicinity of about 0.4 msec, and, that the first few quadrupole 
oscillations would have periods that are also fractions of a millisecond.
Moreover, these oscillation periods were estimated to be damped by 
gravitational radiation with damping timescales of the order of one second.
\par
If an oscillating neutron star is also rotating, then rotation can provide
a coupling between the radial and the nonradial modes, which leads to rapid
loss of radial vibrational energy \cite{whe66}. As pulsars are supposed to be
rotating neutron stars, the study of radial pulsation frequency of rotating
neutron stars constitutes an interesting astrophysical problem to 
investigate. The problem in its details has not been worked out yet in a fully
general relativistic framework with arbitrary rate of rotation, as it is a very
complicated task. In recent years, a Newtonian theory for the above purpose
became available mainly due to the work of Ipser and Managan \cite{ips85} 
Managan \cite{man85} and Ipser and Lindblom \cite{ips89}. 
More recently, a formalism for computing the oscillation
frequencies of rapidly rotating neutron stars is given by Cutler and Lindblom \cite{cut92}
using the post-Newtonian approximation. They employed polytropic equations of 
state, which are an idealization to describe neutron star matter at high
densities and which are unlikely to reflect accurately the adiabatic index 
as function of density, an important ingredient to describe the stability 
of oscillation frequencies. Therefore, it is desirable to use more realistic
equations of state for neutron star matter and a general relativistic 
formalism.
\par
Chandrasekhar and Friedman \cite{cha72} have given a general relativistic formalism
to calculate the effect of rotation ( to order $\Omega^2$, where $\Omega$
is the angular velocity of rotation ) on the eigenfrequencies of radial
pulsations of stars. This formalism is an exact formula to calculate the
frequency ($\sigma^{\prime}$) of oscillations of a `slowly' rotating stellar
configuration which depends only on a knowledge of the Lagrangian displacement
associated with the radial mode of oscillation of the nonrotating 
configurations and of the uniform spherical deformation caused by the rotation.
In this paper, we apply this formalism using recently suggested realistic 
equations of state to estimate the rotationally shifted eigenfrequencies
of radial oscillations of neutron stars.
\par
The Chandrasekhar-Friedman formula is of the form:
\begin{equation}
{\sigma^{\prime}}^2 I_1=I_2+I_3+I_4,
\label{sigmap}
\end{equation}
\par
where
\begin{eqnarray}
I_1=\int\int\int\vec{dx}\sqrt{-g}e^{\lambda}\xi^2 (u^o)^2 (\rho c^2+p)
\label{i1}\\
I_2=\int\int\int\vec{dx}\sqrt{-g}
\left \{ 
\Gamma p 
(\frac{\Delta N}{N})^2 +
(\rho c^2+p)^{-1}\frac{dp}{dr}\frac{d(\rho c^2)}{dr}\xi^2
\right. \nonumber \\
\left.
-2 (\frac{\Delta N}{N})\frac{dp}{dr}\xi-2(p+\frac{1}{8\pi G r^2})(\delta 
\lambda)^2
\right\} \label{i2}\\
I_3=\int\int\int\vec{dx}\sqrt{-g_{(S)}}V^2
\left \{ 
\frac{\Gamma^2 p^2}{\rho c^2+p}
(\frac{\Delta N}{N})^2 + 2\Gamma p (\frac{\Delta N}{N})\frac{d\ln u_1}{dr}\xi
\right. \nonumber \\
\left.
+(\rho c^2+p)(\frac{d}{dr}\ln u_1)^2\xi^2 +2(\frac{\delta V}{V})\delta p
\right. \nonumber \\
\left.
+ 16\pi G(\rho c^2+p)^2e^{\lambda}\xi^2 +2(\rho c^2+p) u^o u_1\frac{d\bar w}{dr}
\frac{\delta\lambda}{V^2}\xi
\right\}_{(S)} \label{i3}\\
I_4=\int\int\int\frac{\vec{dx}}{16\pi G}r^4 sin^3\theta e^{-\lambda/2}
e^{-\nu/2}(\frac{d\bar w}{dr})^2(\delta \lambda)^2
\label{i4}
\end{eqnarray}
\noindent The subscript (S) refers to Schwarzschild (i.e., nonrotating 
configuration) values and $\bar w(r)$ is the angular velocity of the fluid
element relative to the local inertial frame.
\par
In the above equations only the spherical deformation effects of rotation
are manifest. Contributions due to the quadrupole deformation are all zero.
The metric that will correspond to this was given by Hartle \cite{har67} (see also Hartle
and Thorne \cite{har68}
\begin{equation}
ds^2=-e^{\nu}c^2 dt^2 + r^2 sin^2\theta(d\phi-wdt)^2+e^{\lambda}dr^2
+r^2 d\theta^2
\label{metric}
\end{equation}
\noindent where $\omega (r)$ is the angular velocity of the cumulative inertial
frame dragging. In terms of $\omega (r)$, the angular velocity of the local 
inertial frame is written as
\begin{equation}
\bar \omega = \Omega - \omega,
\end{equation}
\noindent where $\Omega$ is the angular velocity as seen by a distant observer.
In Eq.(\ref{metric}), one can write 
\begin{eqnarray}
\nu=\nu_{(S)}+\nu_{(\Omega)},\\
\lambda = \lambda_{(S)}+\lambda_{(\Omega)},
\end{eqnarray}
\noindent to distinguish the corrections of $O(\Omega^2)$ due to the rotation
from the Schwarzschild values by the subscripts ($\Omega$) and ($S$) 
respectively. In Hartle's notation,
\begin{equation}
\nu_{(\Omega)}=h_o
\end{equation}
\noindent and
\begin{equation}
\lambda_{(\Omega)}=\frac{m_o}{r - 2 G m(r)/c^2},
\end{equation}
\noindent where $h_o$ and $m_o$ are functions of $r$. To $O(\Omega^2)$,
\begin{eqnarray}
\sqrt{-g}=r^2 sin \theta[1+(\lambda+\nu)_{(\Omega)}]
{e^{(\lambda+\nu)_{(S)}}},\\
u^o=(1-\nu_{(\Omega)}+\frac{1}{2}V^2)e^{-\nu_{(S)}},\\
V=r\bar \omega sin \theta e^{-\nu_{(S)}},
\end{eqnarray}
\noindent and
\begin{equation}
u_1=r V sin \theta.
\end{equation}
\noindent The various other quantities appearing in Eqs.(\ref{i1} - \ref{i4}) 
are as follows:
\begin{eqnarray}
\delta \lambda = -\frac{4\pi G}{c^2}\frac{r (\rho +p/c^2)}{(1-V^2)}e^{\lambda}
\xi ,\\
(\frac{\Delta N}{N})_{(S)}=\frac{1}{2}\xi\frac{d\nu_{(S)}}{dr}-\frac{d\xi}{dr}
-\frac{2\xi}{r},\\
\frac{\Delta N}{N}=V^2\left\{ \frac{\Gamma p}{\rho c^2+p}(\frac{\Delta N}{N})
+\frac{\xi}{r}\right\}_{(S)} \nonumber\\
-\xi(\frac{2}{r}+\frac{d \lambda}{dr})-\frac{d\xi}{dr}-\delta\lambda,\\
\frac{\delta V}{V}=-(\frac{\Gamma p}{\rho c^2+p}\frac{\Delta N}{N})_{(S)}
-\xi(\frac{2}{r}-\frac{1}{2}\frac{d \nu_{(S)}}{dr}+\frac{1}{\bar\omega}
\frac{d\bar \omega}{dr}).
\end{eqnarray}
\par
The equation governing infinitesimal radial pulsations of a
nonrotating star in general relativity was given by Chandrasekhar \cite{cha64},
and it has the following form :
\begin{equation}
F \frac {d^2\xi}{dr^2} + G \frac{d\xi}{dr} + H\xi = \sigma^2\xi ,
\label{puls}
\end{equation}
\noindent where $\xi$(r) is the Lagrangian fluid displacement and
$c\sigma$ is the characteristic eigenfrequency ($c$ is the velocity
of light). The quantities $F$, $G$,
$H$ depend on the equilibrium profiles of the pressure and density of
the star and are given by
\begin{eqnarray}
F = - e^{-\lambda} e^{\nu} \Gamma p/(p+\rho c^2)
\label{f} ,\\
G = - e^{-\lambda} e^{\nu} \Bigl\{\Gamma p \Big(\frac {1}{2}
\frac{d\nu}{dr} + \frac{1}{2} \frac{d\lambda}{dr} +
\frac{2}{r}\Big) + \nonumber \\
p \frac{d\Gamma}{dr} + \Gamma \frac{dp}{dr}\Bigr\}/(p+\rho c^2),\\
H = \frac {e^{-\lambda}e^{\nu}}{p + \rho c^2} \Bigl\{ \frac{4}{r}
\frac{dp}{dr} - \frac{(dp/dr)^2}{p + \rho c^2} - A \Bigr\} + \frac
{8\pi G}{c^4} e^{\nu} p ,
\end{eqnarray}
\noindent where $\Gamma$ is the adiabatic index, defined in the
general relativistic case as
\begin{equation}
\Gamma = (1 + \rho c^2/p) \frac{dp}{d(\rho c^2)} ,
\end{equation}
\noindent and
\begin{eqnarray}
A = \frac {d\lambda}{dr} \frac{\Gamma p}{r} + \frac
{2p}{r} \frac{d\Gamma}{dr} + \frac {2\Gamma}{r} \frac{dp}{dr} -
\frac{2\Gamma p}{r^2}\nonumber\\
- \frac{1}{4} \frac{d\nu}{dr} \Big( \frac{d\lambda}{dr} \Gamma p +
2p \frac{d\Gamma}{dr} + 2\Gamma \frac{dp}{dr} - \frac{8\Gamma
p}{r}\Big)\nonumber\\ - \frac{1}{2} \Gamma p
\Big(\frac{d\nu}{dr}\Big)^2 - \frac{1}{2} 
\Gamma p \frac{d^2\nu}{dr^2}.
\label{a}
\end{eqnarray}
\noindent 
Note that in Eqs.(\ref{f}-\ref{a}), $p(r),~\rho(r),~\nu(r)$ and $\lambda(r)$
correspond to the values in the nonrotating configuration, whose profiles 
are obtained 
from the equations for the hydrostatic equilibrium of nonrotating degenerate
stars in general relativity \cite{mis70}
\begin{equation}
\frac {dp}{dr} = - \frac{G (\rho + p/c^2) (m + 4\pi r^3 p/c^2)} {r^2
(1-2 Gm/rc^2)} ,
\label{dpdr}
\end{equation}
\begin{equation}
\frac {dm}{dr} = 4\pi r^2\rho ,
\end{equation}
\begin{eqnarray}
\frac{d\nu}{dr} = \frac{2G}{r^2c^2} \frac{(m + 4\pi r^3
P/c^2)}{(1-2Gm/rc^2)},\\
\lambda = -\ln (1-2Gm/r c^2).
\label{lamb}
\end{eqnarray}
\noindent 
The boundary conditions to solve the pulsation equation (\ref{puls}) are
\begin{eqnarray}
\xi (r = 0) = 0,
\label{xi0} \\
\delta p (r = R) = -\xi \frac{dp}{dr} - \Gamma p \frac
{e^{\nu/2}}{r^2} \frac{\partial}{\partial r} (r^2 e ^{-\nu/2}
\xi)\vert_{r=R} = 0
\label{delp}
\end{eqnarray}
\noindent Since $p$ vanishes at $r=R$, it is generally sufficient to demand
\begin{equation}
\xi~~ finite~~ at~~ r=R.
\end{equation}
\noindent Equation (\ref{puls}), subject to the boundary conditions Eq.
(\ref{xi0}) and (\ref{delp}), is a Sturm - Liouville eigenvalue equation 
for $\sigma^2$. 
\noindent The following results follow from the theory of such equations :
\begin{enumerate}
\item The eigenvalues $\sigma^{2}$ are all real.
\item The eigenvalues from an infinite discrete sequence 
\begin{eqnarray}
\sigma^2_o < \sigma^2_1 <\dots < \sigma^2_n <\dots\dots,
\end{eqnarray}
with the corresponding eigenfunctions $\xi_0(r),~\xi_1(r),~...,\xi_n(r)$.
The importance of this result is that 
if the fundamental radial mode of a star is stable ($\sigma_o^2 > 0$)
, then all the radial modes are stable. Conversely, if the star is radially
unstable, the fastest growing instability will be via the fundamental
mode ($\sigma_o^{2}$ more negative than all other $\sigma_n^{2}$).
\item The eigenfunction $\xi_o$ corresponding to $\sigma_o^{2}$ has no
nodes (fundamental mode) in the interval $0<r<R$; more generally,
$\xi_n$ has n nodes in this interval. 
\item  The $\xi_n$ form a complete set for the expansion of any function
satisfying the boundary conditions Eqs. (\ref{xi0}) and (\ref{delp}).
\end{enumerate}
\noindent 
Given an equation of state p($\rho$), Eqs. (\ref{dpdr}) and (\ref{lamb}), called
Tolman-Oppenheimer-Volkoff (TOV) equation, can be numerically integrated for a
given central density to obtain the radius $R$ and gravitational mass $M =
m(R)$ of the star. 
\par
For the rotational spacetime corresponding to the metric Eq.(\ref{metric}), 
the stellar structure equations (\ref{dpdr}) -(\ref{lamb}) must be 
supplemented by additional
equations that provide the profiles for $\bar \omega(r)$ and the (rotation 
induced) pressure and mass deformation terms. The details of which are given
in references \cite{har67} and \cite{har68}, and so are not reproduced here. For the integrand 
appearing in Eq.(\ref{i3}), the subscript (S) refers to the Schwarzschild values
of quantities $p,~\rho$ etc. In the Eqs.(\ref{i1}), (\ref{i2}) and (\ref{i4}) 
all the
quantities must incorporate the rotational contributions.
\par
The main ingredient to solve for the structure and pulsation features is the 
equation of state (EOS), $p=p(\rho)$. The range of $\rho$ expected for neutron 
stars spans a very wide range: from about $7.8~g~cm^{-3}$ near the surface to
$\rho > 10\rho_o$ in the interior, where $\rho_o$ is the equilibrium nuclear
matter density (=2.8 $\times 10^{14}~g~cm^{-3}$). Structure parameters of a
neutron star depend sensitively on the EOS at high densities \cite{arn77,
dat89} especially
around the density region $10\rho_o$. Therefore, the oscillation features
are also expected to possess marked sensitivity on the high density EOS,
and so it becomes imperative to use realistic neutron star EOS model rather
than a polytropic form for the EOS. 
\par
Despite two decades of theoretical research, there exists no clear consensus
on the behaviour of the EOS at high densities. The nearly two dozen EOS 
models available in the literature comprise a rather broad set. For our purpose
here, we adopt a realistic non-relativistic model given by Wiringa, 
Fiks and Fabrocini \cite{wir88} and a relativistic model given by Sahu, Basu
and Datta \cite{sah93}. The non-relativistic
model describes the high density matter in the interior region of neutron star
as beta-stable neutron matter. This model possesses desirable properties for
two-nucleon scattering phase shifts, equilibrium nuclear matter and binding
energies of light nuclei. The three versions are given by Wiringa, Fiks and
Fabrocini \cite{wir88} which correspond to a variation in the choice of 
the nuclear 
interaction involved. We choose here the UV14+UVII (neutron, proton, electrons
and muons). Whereas, in recent times, the relativistic model has drawn 
considerable attention. In the relativistic approach, one
usually starts from a local, renormalizable field theory with
baryon and explicit meson degrees of freedom. The theory is
chosen to be renormalizable in order to fix the coupling
constants and the mass parameters by empirical properties of
nuclear matter at saturation. As a starting point, one chooses
the mean field approximation which should be reasonably good at
very high densities. This approach is currently used as a
reasonable way of parameterizing the EOS. However, in recent
years, the importance of the three-body forces in the EOS at
high densities has been emphasized by several authors (Jackson, Rho
and Krotscheck \cite{jac85}; Ainsworth {\it et al.} \cite{ain87}).
This gives theoretical impetus to study the chiral sigma model,
because the non-linear terms in the chiral sigma Lagrangian can
give rise to the three-body forces. In the present calculation, we take 
the SU(2) chiral sigma model
EOS for the neutron-rich matter in beta equilibrium \cite{sah93}.
The composite equations of state to determine the neutron star 
structure was set up as follows. For the region $3.7\times 10^{11} < \rho <
10^{14} g~cm^{-3}$, $p(\rho)$ is taken from Negele and Vautherin \cite{neg73}, 
while
for $\rho < 3.7\times 10^{11} g~cm^{-3}$, $p(\rho)$ is taken as given by Baym,
Pethick and Sutherland \cite{bay71} and Feynman, Metropolis and Teller \cite{fey49}.
\par
We first solved the stellar structure equations that provide the nonrotating
neutron star configuration. The numerical integration was done using the
modified Euler method with variable step size (where density gradient is large)
and a logarithimc interpolation for the ($p,~\rho$) data. The method was tested
by requiring agreement with the results for the neutron star mass, radius and
momemt of inertia as reported by Wiringa, Fiks and Fabrocini \cite{wir88}. The rotational
contribution to $p,~ \rho,~ m$ and $\nu$ were calculated similarly (to accuracy
$<0.6\%$, compared with the published results by Hartle and Thorne \cite{har68} for the
case of Harrison-Walker-Wheeler EOS). The nonrotating general relativistic
pulsation Eq.(\ref{puls}) was replaced by a set of difference equations which were
cast in tridiagonal form. The eigenvalue and eigenfunction were evaluated using
EISPACK routine. The eigenvalues calculated by us agree with the radial 
pulsation frequencies of nonrotating neutron stars for the Wiringa, Fiks and
Fabrocini \cite{wir88} EOS calculated by Cutler, Lindblom abd Splinter \cite{cut90}. Finally, the
integrals $I_1,~ I_2,~ I_3$ and $I_4$ were evaluated using standard integration
package, and then substituted in Eq.(\ref{sigmap}) to obtain the value of 
$\sigma^{\prime}$, the eigenfrequency of raidial pulsation of the star in
rotation. This was done for several values of the rotation rate ($\Omega$)
to obtain $\sigma^{\prime}(\Omega)$. For purpose of illustration, we consider
only one configuration, namely a $1.4~M_{\odot}$ ($M_{\odot}=$ solar mass)
neutron star. Masses of neutron stars estimated from binary pulsar data are
all consistent with a value 1.4 $M_{\odot}$.
\par
The results of our calculation are presented in figures 1 and 2 and table 1. In 
figure 1, we show the range in the values of the frequency of radial 
oscillation for the nonrotating neutron stars configurations. In this figure,
$P$ stands for the period of oscillation and is plotted as a function of the
stable gravitational mass of the nonrotating neutron stars (with maximum
stable mass equal to 2.59 $M_{\odot}$ for EOS \cite{sah93} in Fig. 1(a) 
and 2.18 $M_{\odot}$
for EOS \cite{wir88} in Fig. 1(b)); curve 1 refers to the fundamental
mode of oscillation and the curves 2 - 5 to the first four harmonics. For the
fundamental mode, we find that $P$ lies in a rather narrow range of values:
(0.3-0.5)msec for both the EOS. For the higher modes considered in 
Fig. 1(a) and 1(b), values of $P$ lie roughly between the limits 0.25 msec 
and 0.1 msec, and seem to saturate at about 0.1 msec. In case of quark 
stars \cite{dat92},
the fundamental mode of oscillation periods are in a range of values 
(0.05 - 0.3) msec and for higher modes, the periods are $\le 0.1$ msec. 
Thus, the neutron star periods are slightly higher than for quark stars.
The pulsation frequency
as a function of $\Omega/\Omega_c$ for the fundamental model (n=0) for both
relativistic EOS \cite{sah93} and non-relativistic EOS \cite{wir88} are
shown in Fig. 2(a) and 2(b) respectively. In both the cases, it is found that
the pulsation frequency is an increasing function of $\Omega$. In Fig. 2(b),
rate of increase is higher, which is possibly due to the nature of interaction 
used in EOS. The EOS is stiff for relativistic case \cite{sah93}, whereas, it
is soft in the non-relativistic case \cite{wir88}. However, we do not find such
a feature in the nonrotating case.
Table 1 presents the values of $\sigma^{\prime}(\Omega)$
for the fundamental mode and the first two harmonics for both relativistic
EOS \cite{sah93} (a) and non-relativistic EOS \cite{wir88} (b). 
For comparison, we
have included in this table the values of $\sigma$, the pulsation frequency
of the nonrotating stars. The parameters appearing in this table are all for
a 1.4 $M_{\odot}$ neutron star. We have chosen the rotation rate $\Omega$
to vary from 0.1$\Omega_c$ to 0.5$\Omega_c$, where $\Omega_c=(GM/R^3)^{1/2}$.
The upper limit 0.5$\Omega_c$ is almost equal to the termination point
0.6$\Omega_o$, where $\Omega_o=(3GM/4R^3)^{1/2}$, for sequences of rotating
stellar models, applicable to both Newtonian and general relativistic
stellar models and is independent of the equation of state of the stellar 
matter \cite{fri86}. We have, however, not investigated the dependence of this 
termination limit on the pulsation degrees of freedom.
\par
From table 1,  the following behaviour for $\sigma^{\prime}(\Omega)$ emerges.
For the fundamental mode (n=0), $\sigma^{\prime}$ possess a monotonic 
dependence on $\Omega$ and it is found to be always an increasing function
of $\Omega$. The same feature is also found for n=1 and n=2 modes for both
EOS (a) and (b). In EOS (a), the $\sigma^{\prime}$ is very 
sensitive to $\Omega$ for n=0 and n=1. This is due to the nature of force 
involved in EOS. On the otherhand, $\sigma^{\prime}$ seems to saturate for
n=2 in EOS (a) and n=0 and n=1 in EOS (b).
\par
We have thus applied the general relativistic formalism by Chandrasekhar 
and Friedman \cite{cha72} to calculate frequencies of radial pulsations
of rotating neutron stars, for realistic EOS. However, the formalism is based
on the use of Hartle's prescription \cite{har67} to treat the rotation general
relativistically (to order $\Omega^2/{\Omega_c}^2$). The essential physics
of high density matter through the incorporation of a realistic EOS is
important, and this feature is present in our work. The main conclusion 
of our work is that the magnitude of the effect of the dynamical coupling
of oscillation and rotation will differ substantially in a general 
relativistic treatment of the problem as compared to Newtonian and post-
Newtonian approximation approaches. Extension of this work to include the
damping time scale with other different EOS will be reported in a future
work.
\par
It is great pleasure to thank B. Datta and S. S. Hasan from Indian
Institute of Astrophysics, Bangalore, for useful and fruitful discussions 
during the preliminary stage of this work.

\newpage
\begin{table}
\caption { Pulsation frequency as a function of the rotation rate, 
$\sigma^{\prime}(\Omega)$, for n=0, 1 and 2 radial modes. 
$\sigma$ stands for the radial
pulsation frequency of the nonrotating neutron stars. The neutron stars 
configuration  correspond to a mass of 1.4$M_{\odot}$ with: (a) relativistic
EOS [19] and (b) non-relativistic EOS [18].}
\vskip 0.2 in
\begin{tabular}{ccccccccc}
\hline
\multicolumn{1}{c}{$n$} &
\multicolumn{1}{c}{$\Omega/\Omega_c$}&
\multicolumn{1}{c}{~~~~~~~~~~~~(a)} &
\multicolumn{1}{c}{} &
\multicolumn{1}{c}{~~~~~~~~~~~~(b)} &
\multicolumn{1}{c}{} \\
\hline
\multicolumn{1}{c}{} &
\multicolumn{1}{c}{} &
\multicolumn{1}{c}{$\sigma$}&
\multicolumn{1}{c}{$\sigma^{\prime}(\Omega/\Omega_c)$}&
\multicolumn{1}{c}{$\sigma$}&
\multicolumn{1}{c}{$\sigma^{\prime}(\Omega/\Omega_c)$}\\
\multicolumn{1}{c}{} &
\multicolumn{1}{c}{} &
\multicolumn{1}{c}{($10^{4}s^{-1}$)} &
\multicolumn{1}{c}{($10^{4}s^{-1}$)}&
\multicolumn{1}{c}{($10^{4}s^{-1}$)} &
\multicolumn{1}{c}{($10^{4}s^{-1}$)}\\
\hline
0&0.1&2.108&5.614&2.066&2.081\\
 &0.2&     &6.625&     &2.123\\
 &0.3&     &8.654&     &2.184\\
 &0.4&     &10.464&    &2.251\\     
 &0.5&     &11.839&    &2.313\\
\hline
1&0.1&3.666&6.598&4.979&5.012\\
 &0.2&     &7.529&     &5.102\\
 &0.3&     &9.062&     &5.222\\
 &0.4&     &10.536&    &5.340\\
 &0.5&     &11.396&    &5.436\\
\hline
2&0.1&5.379&5.721&7.155&7.400\\
 &0.2&     &6.648&     &7.963\\
 &0.3&     &6.540&     &8.545\\
 &0.4&     &6.776&     &8.975\\
 &0.5&     &6.847&     &9.213\\
\hline
\end{tabular}
\end{table}
\newpage
\begin{figure}
\caption {Periods ($P$) of radial oscillations versus stable gravitational
mass of nonrotating neutron stars: (a) relativistic EOS [19] and
non-relativistic EOS [18]. Curve 1 is
for the fundamental mode and curves 2 - 5 correspond to the first four
harmonics.}
\end{figure}
\begin{figure}
\caption {The pulsation frequency ($\sigma^{\prime}$) as a function of 
$\Omega/\Omega_c$ of rotating neutron stars: (a) relativistic 
EOS [19] and non-relativistic EOS [18].}
\end{figure}
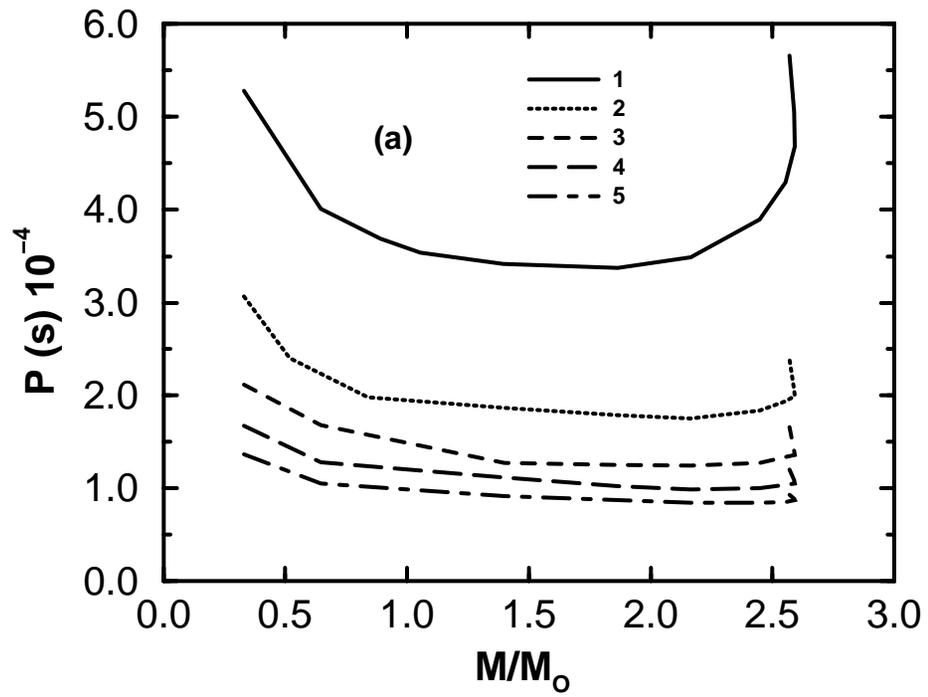
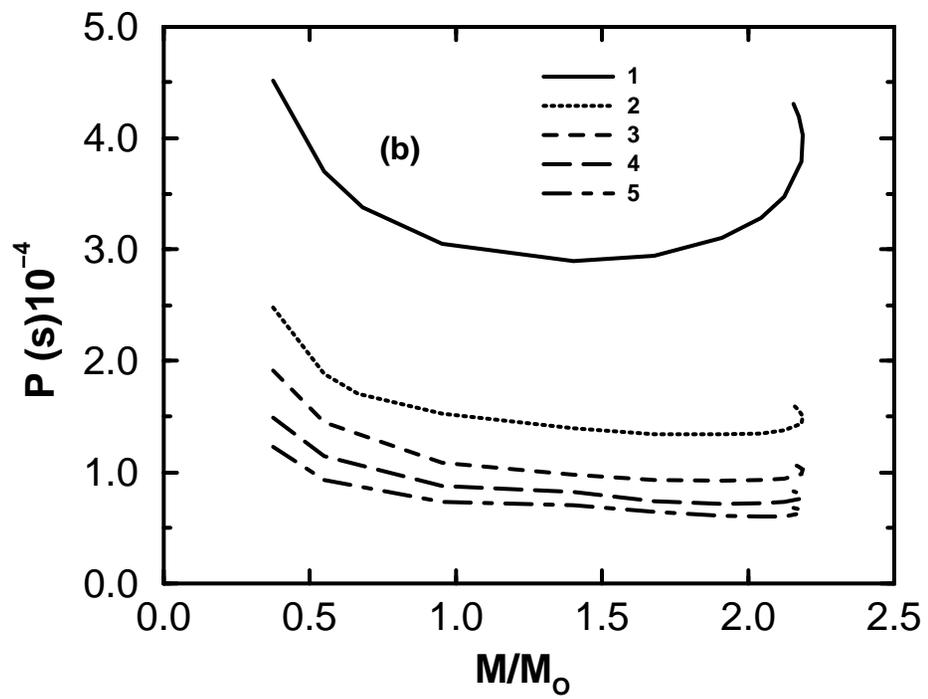
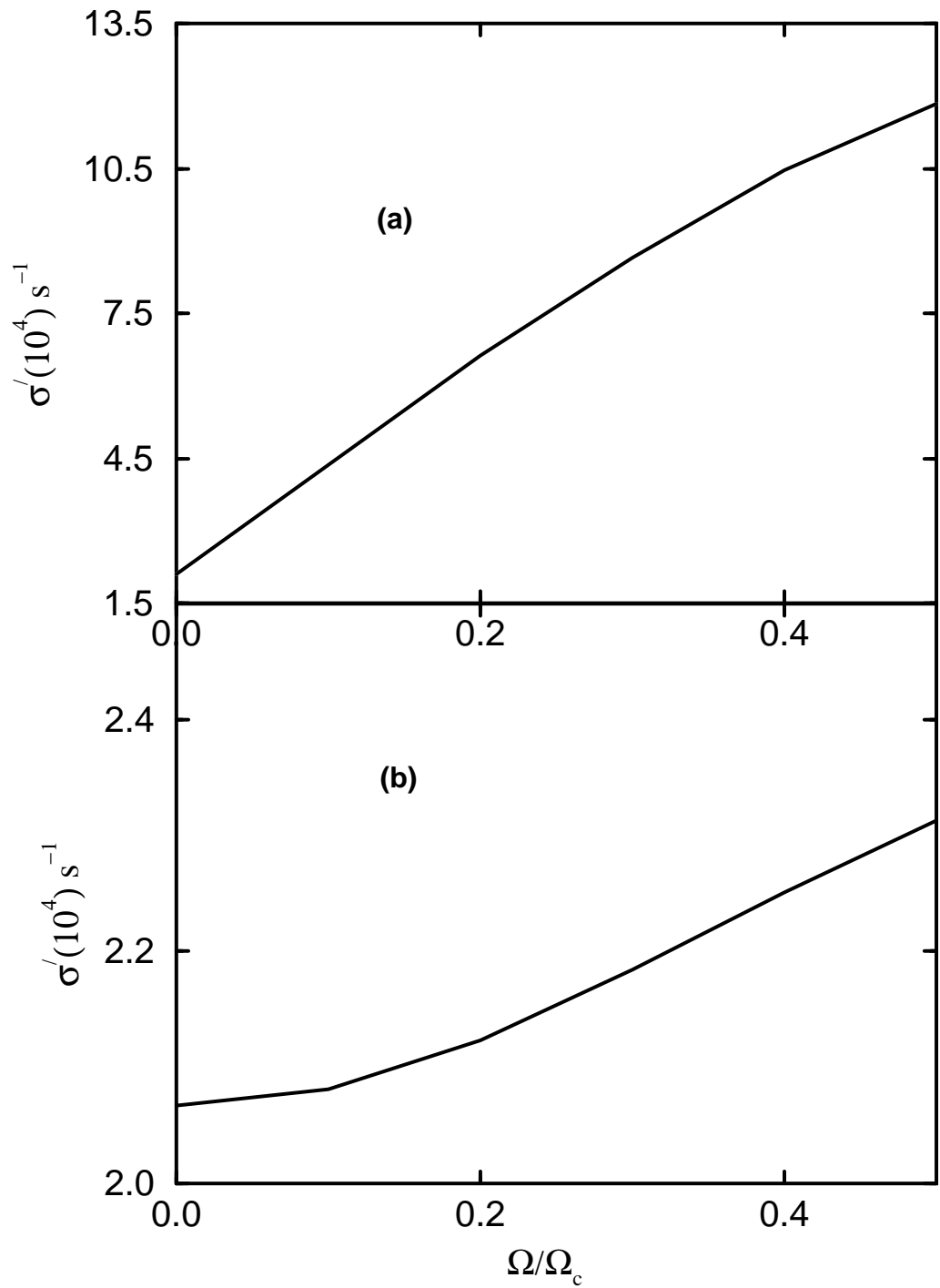
\end{document}